\documentclass[preprint]{aastex}
\shorttitle{Slipping magnetic reconnections with multiple flare ribbons}
\slugcomment{Submitted to ApJ}
\shortauthors{Zheng et al.}
\begin{document}

\title{Slipping magnetic reconnections with multiple flare ribbons during an X-class solar flare}
\author{Ruisheng Zheng, Yao Chen, and Bing Wang}
\affil{Shandong Provincial Key Laboratory of Optical Astronomy and Solar-Terrestrial Environment, and Institute of Space Sciences, Shandong University, 264209 Weihai, China; ruishengzheng@sdu.edu.cn}

\begin{abstract}
With the observations of the Solar Dynamics Observatory, we present the slipping magnetic reconnections with multiple flare ribbons (FRs) during an X1.2 eruptive flare on 2014 January 7. A center negative polarity was surrounded by several positive ones, and there appeared three FRs. The three FRs showed apparent slipping motions, and hook structures formed at their ends. Due to the moving footpoints of the erupting structures, one tight semi-circular hook disappeared after the slippage along its inner and outer edge, and coronal dimmings formed within the hook. The east hook also faded as a result of the magnetic reconnection between the arcades of a remote filament and a hot loop that was impulsively heated by the under flare loops. Our results are accordant with the slipping magnetic reconnection regime in 3D standard model for eruptive flares. We suggest that complex structures of the flare is likely a consequence of the more complex flux distribution in the photosphere, and the eruption involves at least two magnetic reconnections.

\end{abstract}

\keywords{magnetic reconnection --- Sun: activity --- Sun: corona}

\section{Introduction}
Solar flares are most energetic magnetic explosions in the solar activities. They can increase the emission in a broad range of the electromagnetic spectrum, from radio wavelengths to X- and $\gamma$-rays (Fletcher et al. 2011). In the standard solar flare model, i.e. the CSHKP model (named after Carmichael 1964; Sturrock 1966; Hirayama 1974; Kopp \& Pneuman 1976), the erupting flux rope stretched magnetic filed lines to induce the magnetic reconnection; due to the successive reconnections, the flare loops (FLs; originally named as post-flare loops) formed and straddled the magnetic polarity inversion line, and their footpoints are heated by the energy transport from the reconnection site to appear as flare ribbons (FRs).

However, the standard flare model is basically two-dimensional, and it remains deficient to explain many inherent three-dimensional (3D) observational features, such as the formation of coronal sigmoids (Aulanier et al. 2010; Green et al. 2011; Savcheva et al. 2015), the erupting flux rope (Zhang et al. 2012), the moving bright emissions along the FRs (Fletcher \& Hudson 2002; del Zanna et al. 2006; Chandra et al. 2009), and the strong-to-weak shear transition in FLs (Aulanier et al. 2012). Recently, Aulanier et al. (2012) and Janvier et al. (2013) proposed a fully 3D flare model that incorporates the standard 2D one in one of its cuts, thus satisfying the principle of correspondence. In the 3D magnetohydrodynamic simulation, the magnetic field lines passed through the quasi-separatrix layers (QSLs; Priest \& D\'{e}moulin 1995), and could undergo a successive reconnection, which exchanged their connectivity with neighboring field lines; the continuous reconnections resulted in the obvious slipping motion along the QSLs (Pontin et al. 2005; De Moortel \& Galsgaard 2006; Aulanier et al. 2006). Therefore, it is classified as the slipping magnetic reconnection.

As the theoretical models have been developed, some observational cases of the slipping magnetic reconnection have been analysed. Aulanier et al. (2007) first presented the direct observations of slipping magnetic reconnection in coronal loops by the X-ray Telescope onboard Hinode. Based on the Atmospheric Imaging Assembly (AIA; Lemen et al. 2012) on the Solar Dynamics Observatory (SDO; Pesnell et al. 2012), Dud\'{i}k et al. (2014) reported the apparent slipping motion of FLs during an eruptive X1.4 flare, and Li \& Zhang (2014, 2015) showed two examples of slipping motion of FLs and the quasi-periodic pattern of the latter case. However, the previous observations only focused on the slipping motion in two FRs, and the slippage in multiple FRs with complex magnetic configuration has never been discussed. In this Letter, we present the slipping magnetic reconnections in multiple FRs during an X1.2 eruptive flare.

\section{Observations and Data Analysis}
The X1.2 flare occurred in ARs 11944 and 11943 on 2014 January 7, associated with a rapid coronal mass ejection and a fast coronal wave. We only focus on the flare, and its start, peak, and end times are about 18:04, 18:32, and 18:58 UT, respectively. To study the evolution of FRs, we mainly use the observations from the AIA on the SDO and the H$\alpha$ filtergrams from the Global Oscillation Network Group (GONG) of the National Solar Observatory. The AIA has 10 EUV and UV wavelengths, covering a wide range of temperatures. The AIA image (4096~$\times$4096 pixels) covers the full disk of the Sun and up to 0.5 $R_\odot$ above the limb, with a pixel resolution of 0.6$"$ and a cadence of 12 s. The H$\alpha$ images are at 6563~{\AA} with a spatial resolution of 1$"$ and a cadence of around 1 minute (Harvey et al. 2011). Magnetograms from the Helioseismic and Magnetic Imager (HMI; Scherrer et al. 2012), another instrument on the SDO, are chosen to check the magnetic field configuration of the eruption region, with a cadence of 45 s and a pixel scale of 0.6$"$. In order to analyze the dynamics of moving structures, we employ the time-slice approach. The associated speeds are obtained by the linear fits, assuming that the measurement uncertainty of the selected points is 4 pixels ($\sim1.74$ Mm).

\section{Results}
The general appearance of the eruption region and the nearby environment before the flare is shown in the HMI magnetogram (panel a) and AIA 171~{\AA} image (panel b) in Figure 1. The flare occurred between the ARs 11944 and 11943, but the negative polarity (N1) of AR 11944 barely contributed. The flare was associated with the leading positive polarity (P1) of AR 11944 and the pair of polarities of AR 11943 (N2-P2), and it seems that N2 is surrounded by P1 and P2 (panel a). There was a sigmoidal loop (SL) connecting N2 and P1 (panels b-c). To the north of SL, there were trans-equatorial lines (TELs) linking ARs 11944 and 11946. TELs were resulted from magnetic reconnection that occurred at the X point between ARs 11944 and 11946 (panel b). As TELs formed more and more, their south footpoints approached much more close to SL (boxes in panels d-e). The time-slice plot along the S1 in panel e shows the clear inflow for TELs, with a speed of 4.8 km s$^{-1}$ before the flare onset (panel i), which is close to the upper limit of reconnection inflow speeds (Yokoyama et al. 2001). The close relationship between the inflow and the flare onset demonstrates that the magnetic reconnection between TELs and the overlying field of SL's western part invoked the flare. There formed three normal FRs (panel f): the center FR (FR1), the east FR (FR2), and the west FR (FR3). FR1 located at N2, and FR2 and FR3 lied at P1 and P2, respectively. FR2 and FR3 connected FR1 by FLs (black arrows in panel h). FRs all experienced a lateral extension, as a result of slipping reconnections. There appeared hook structures at the eastern end of FR1 and FR2 (black arrows in panel g), and a faint semi-circular secondary FR (SFR; named by Zhang et al. 2014) on P2 (the white arrow in panel g) connecting FR3. Note that the warm coronal loops (L1; the white arrow in panel h), connecting SFR and FR1, survived during the eruptive flare.

The evolution of the multiple FRs is shown in 304~{\AA} images (first and third columns) and H$\alpha$ filtergrams (second and fourth columns) in Figure 2. Before the flare, FR3 and SFR were at a single continuous bright lane (white arrows in panel a1). There were also some filaments (F1-F3; arrows in panel a2) around the eruption region. At the beginning of the flare, FR1 lay at the edge of N2 as the coast-line of a peninsula (the black arrow in panel b1); sequentially, FR1 quickly moved from the north top toward the center of N2 (the dotted line S2 and the southward black arrow in panel c1). Meanwhile, FR2 extended eastward, and FR3 moved southward (white arrows in panels b1-d1). As a result of the extension, hook-like structures formed at the east ends of FR1-FR2 (conjugated arrows in panel d1), and there appeared the semi-circular SFR connecting the south end of FR3. In addition, there appeared mass flow from FR2 to AR 11946 (black arrows in panels e-f), which likely provides another evidence of the magnetic reconnection between TELs and the overlying field of SL's western part. Moreover, the F1-F3 seemed to survive during the entire flare (white arrows in panels e2-h2).

The attractive element of the event is that the bright features moving along SFR (better seen in the attached animation). The bright features first moved clockwise in west-east direction towards SFR's easternmost point (northward black arrows in panels c-d of Figure 2), and then slipped anticlockwise towards FR3. Note that the slippage of SFR first occurred along the outer edge, and then moved along the inner edge. Following the slippage, SFR nearly disappeared and was replaced by wider coronal dimmings (northward white arrows in panels e1-h1 of Figure 2), but FR3 were nearly intact (black arrows in panels g2-h2 of Figure 2). The slipping motion along SFR and the associated dimmings are consistent with the presence of eruption as predicted by the standard 3D model for eruptive flares (Aulanier et al. 2012; Janvier et al. 2013). Next, we focus on the interesting anticlockwise slippage of SFR shown in AIA 94, 193, 171, 304, and 1600~{\AA} (Figure 3). In the right column, the moving features are indicated by black arrows, and the wider dimmings are pointed out by white arrows in panels a2-d4. The hook-like structures of FR1 and FR2 are shown by black arrows in the left column. During the flare, L1 connecting SFR and N2 (white arrows in panels b1-c1) were intact.

To best analyse the kinetics of FR1 and SFR, we employ the time-slice approach along the dotted lines in Figure 2 and Figure 3. According to the time-slice plots (upper panels of Figure 4), the slippage of FR1 started at about 18:04 UT, close to the flare onset, and its speed is about 384 km s$^{-1}$. The anticlockwise propagation velocity of the moving feature along SFR is about 110-130 km s$^{-1}$ in 304 and 94~{\AA}, and its onset is about at 18:30 UT.

In middle and bottom panels of Figure 4, the eruptive flare is also shown in original (panels d and i) and base-difference (panels e-h) images of AIA 94~{\AA} (see the attached animation). The moving features of SFR (white arrows in panels f-h) and associated dimmings (the white arrow in panel i) are clear. At the beginning of the flare, FLs were still in a lower altitude, but there were some hot loops (L2, panel d) overlying FLs between FR1 and FR2, which existed before the flare. As FLs grew up, L2 was impulsively heated and expanded. About some minutes later, there formed some new loops (L3; the black arrow in panel f) connecting FR1 and a remote brightening (the white arrow in panel e). The brightening rooted in the negative polarity (N3, see Figure 1a) of F3, consistent with the brightening at the turn of F3 (westward white arrows in panels e1-f1 of Figure 2). Because L3 connected P1 with N3, there is likely a magnetic reconnection between L2 and the overlying arcades of F3. L3 kept extending southward along N3 (the black arrow in panels h-i), and there left a clear channel between L3 and FLs. On the other hand, the hook-like structure of FR2 was also fading (dotted boxes in panels e-g), which is possibly related to the formation of L3.


\section{Discussion and Conclusions}
The apparent slipping motion of the FLs and the hook-like structures of FRs are consistent with the 3D reconnection in QSLs, as predicted by the 3D standard flare model (Aulanier et al. 2012 and Janvier et al. 2013). The 3D model is generically with two polarities and two J-shape FRs. This kind of slippage of twin FRs have been reported in previous studies (Dud\'{i}k et al. 2014; Li \& Zhang 2014, 2015). Here, the slipping motions of the flare involved at least two ARs and three distinct FRs, which may be a consequence of the more complex flux distribution in the photosphere. The semi-circular SFR contributes to complex magnetic configuration, and confirms the presence of a QSL (Aulanier et al. 2007).

The SFR was not connected by FLs, and disappeared following the slippage. Does it mean that SFR is a different structure, e.g. footpoints of coronal loops impulsively heated by the flare (Zheng et al. 2015)? Note that the eruption was launched between FR1 and FR3 towards southwest (better seen in 193~{\AA} of the attached animation), and it was closely associated with a fast partial-halo CME in the southwest {\footnote{\url{http://cdaw.gsfc.nasa.gov/movie/make\_javamovie.php?\\
stime=20140107\_1711\&etime=20140107\_2049\&img1=lasc2rdf\&title=20140107.182405.p231g;V=1830km/s}}}. Following the slippage, SFR widened, and coronal dimmings formed within the widened part of SFR (white arrows in right column of Figure 3), which is in accordance with the predictions of the 3D standard model for eruptive flares (Figure 11a of Janvier et al. 2015). Hence, SFR is the extended hook of FR3, and represents footpoints of erupting structures as predicted in the 3D standard model for eruptive flares. The presence of dimmings within the hook is due to the stretching of the field lines opened by the eruption.

According to 3D slipping reconnection in QSLs predicted by the 3D flare model, we suggest the scenario of the slippage of SFR in a sketch (Figure 5), which is based on an HMI magnetogram showing the associated magnetic polarities (P1-P2, N1-N3). FR3 and SFR are at a single continuous bright lane (orange lines) before the flare (Figure 2a1). The erupting SL (the dashed light-blue line) induced the flare and FLs (yellow lines). Due to the successive reconnections, FR3 extended clockwise and formed a semi-circular SFR. The bright features first moved along the outer edge of SFR toward its easternmost point, and continually slipped anticlockwise along the inner edge of SFR, followed by coronal dimmings within the hook (the shadow in panel b). The slipping directions of SFR is indicated by orange arrows, and the tight narrow hook is similar to that reported by Dud\'{i}k et al. (2014). The presence of dimmings reveals the open of the erupting structure (the pink lines). Note that the erupting structure is just a schematic representation of the eruption, and can itself be fragmented, since all of the FR1-FR3 have hook structures. In addition, the survived warm coronal loops (L1; red lines) connected the inner edge of SFR and FR2. Due to the eruption direction, the dimmings only occurred to the southwest of the closed L1. The inner edge of SFR is then simply the topological boundary between closed and open field lines. On the other hand, the hot coronal loops L2 (green lines) are impulsively heated, expanded, and reconnected with the magnetic field lines (black lines) overlying F3 (the sparkle), which resulted in the formation of L3 (deep-blue lines).

As the findings in Dud\'{i}k et al. (2014), the tight narrow hook of SFR fade and disappear, due to the moving footpoints of erupting structures. For the fading and disappearance of the hook of FR2, it is possibly because of the successive reconnections between L2 and the overlying loops of F3. It suggests that the eruption with complex magnetic configurations involves at least two magnetic reconnections.

All the FRs underwent slipping motions in most of AIA passbands, especially for FR1 and SFR. FR1 extended fast with a speed of 384 km s$^{-1}$, which is faster than that of previous results; the anticlockwise propagation velocity of SFR is about 130 km s$^{-1}$, which is similar to that of the previous results (Dud\'{i}k et al. 2014; Li \& Zhang 2014, 2015). The fast slippage of FR1 reveals the strong-to-weak shear transition of FLs, which is likely associated with the untwisting of SL during the eruption. According to the definitions in Aulanier et al. (2006), the slipping-running and slipping reconnection regimes respectively correspond to super- and sub-Alfv{\'e}nic field line fast slippage. Due to the average Alfv{\'e}n speed for the decaying AR is 500 km s$^{-1}$ (Aulanier et al. 2012), all the slipping speeds here are in the range of sub-Alfv{\'e}nic speed. We suggest that the extension and propagation in multiple FRs are as a result of slipping magnetic reconnections.

Benefiting from the high-resolution observations of AIA/SDO, we capture the slipping magnetic reconnections in multiple FRs during the X1.2 flare on 2014 January 7. As the result of slipping motion, multiple FRs extended with hook structures. Due to the moving footpoints of erupting structures and another magnetic reconnection, hooks disappeared, and/or was replaced by coronal dimmings. Our results are accordant with the slipping magnetic reconnection regime in 3D standard model for eruptive flares (Aulanier et al. 2006; Aulanier et al. 2012; Janvier et al. 2013; Janvier et al. 2015). More studies of the slipping reconnection will be helpful in understanding the 3D standard flare model; further observations and theoretical work will be necessary.

\acknowledgments
Many thanks to Wei Liu, Ting Li, Xiaoli Yan, and Hongqiang Song for the constructive discussion. SDO is a mission of NASA's Living With a Star Program. The authors thank the SDO team for providing the data. This work is supported by grants NSBRSF 2012CB825601, NNSFC 41274175, and 41331068, and Yunnan Province Natural Science Foundation 2013FB085.

\clearpage

\begin{figure}
\epsscale{0.9}
\plotone{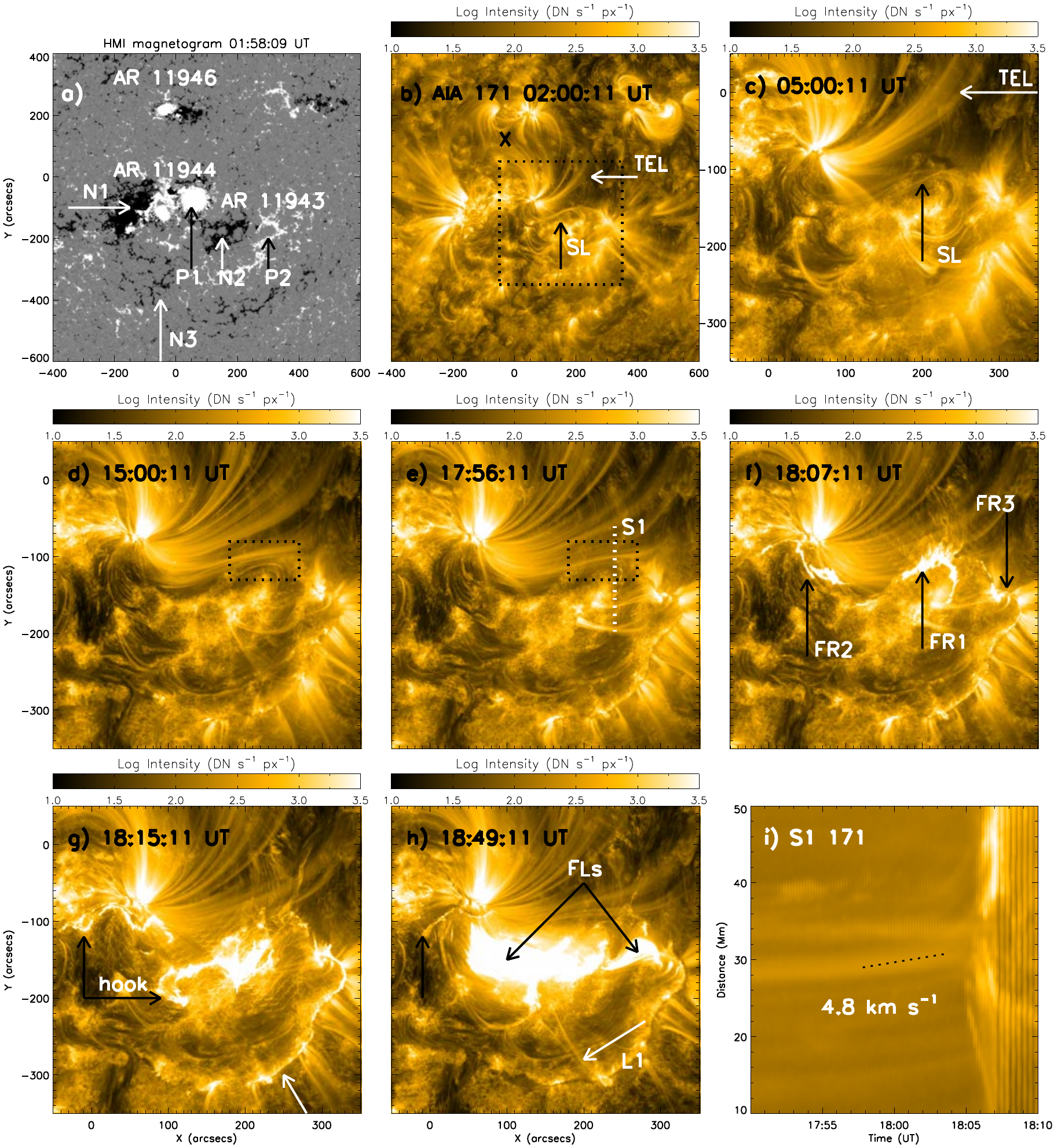}
\caption{The magnetic field environment of the eruption region in an HMI magnetogram (panel a) and the evolution of the flare in AIA 171 ~{\AA} images (panels b-h). The magnetic polarities (P1-P2 and N1-N3) are indicated by the arrows in panel a. The FOV of panels c-h is illustrated by the dotted box in panel b, and the X represents the location of the magnetic reconnection between ARs 11944 and 11946, where TELs formed. The arrows in panels b-c point out the loops (SL and TELs) associated with the magnetic reconnection triggering the flare, and the dotted boxes in panels d-e present the place of this magnetic reconnection. The arrows in panels f-h show the structures (FRs, FLs, and hooks) of the flare, and the survived warm loops (L1). The inflow of the flare is obvious in the time-slice image (panel i) along the dotted lines in panel e.
\label{f1}}
\end{figure}

\clearpage

\begin{figure}
\epsscale{0.9}
\plotone{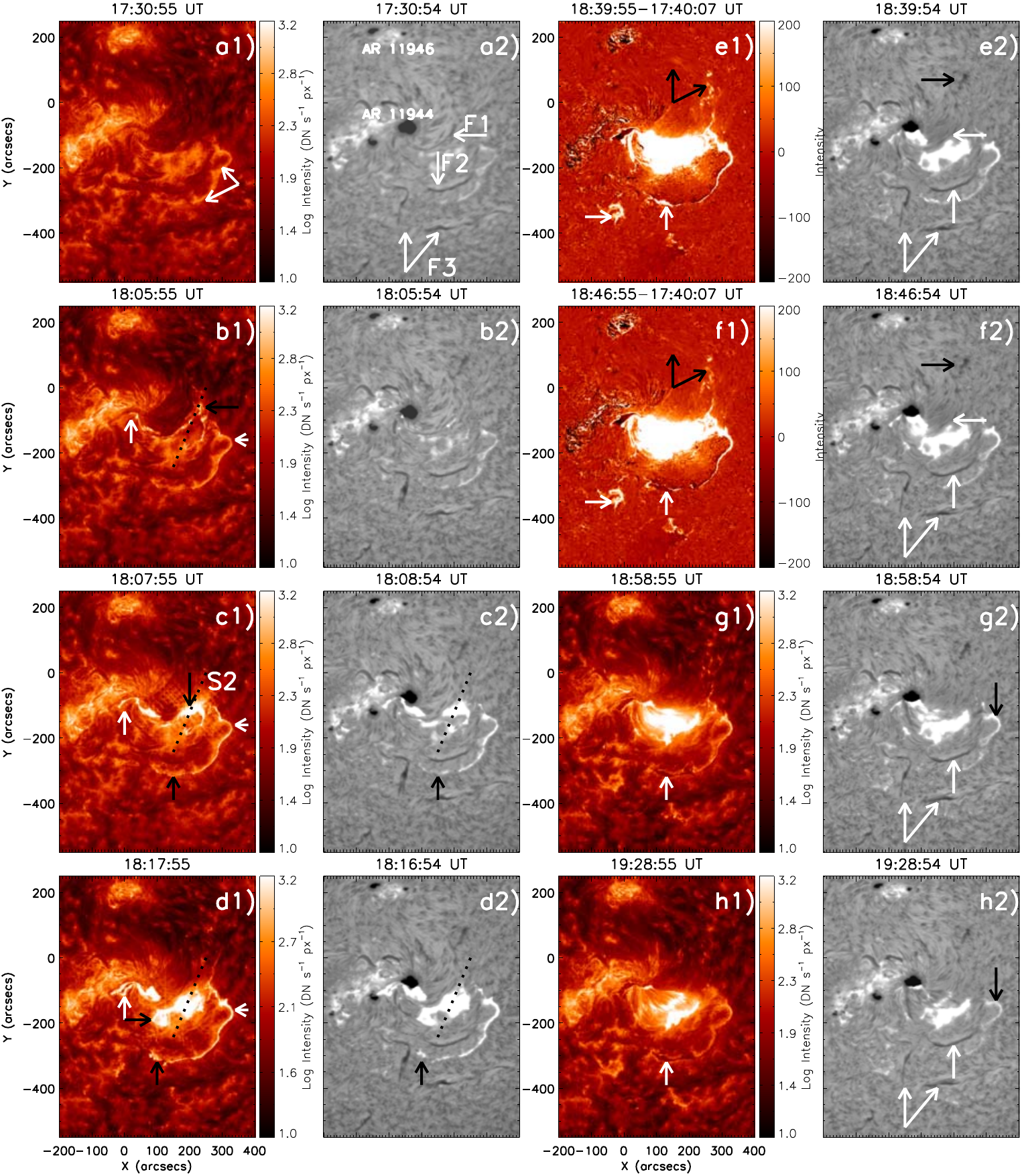}
\caption{The evolution of the multiple FRs in AIA 304~{\AA} (first and third columns) and H$\alpha$ filtergrams (second and fourth columns). Except the westward black arrows in base-difference images (panels e1-f1) showing the brightening, the arrows in panels a1-d1 indicate the multiple FRs, and the white and black arrows in panels e1-h1 point out the dimmings and the mass flow, respectively. In H$\alpha$ filtergrams, the white arrows indicate the filaments (F1-F3), and the black arrows in panels e2-f2 and those in other panels show the mass flow and FRs, respectively. The dotted lines in panels c-d are used to obtain the time-slice image in Figure 4a.
\label{f2}}
\end{figure}

\clearpage

\begin{figure}
\epsscale{0.6}
\plotone{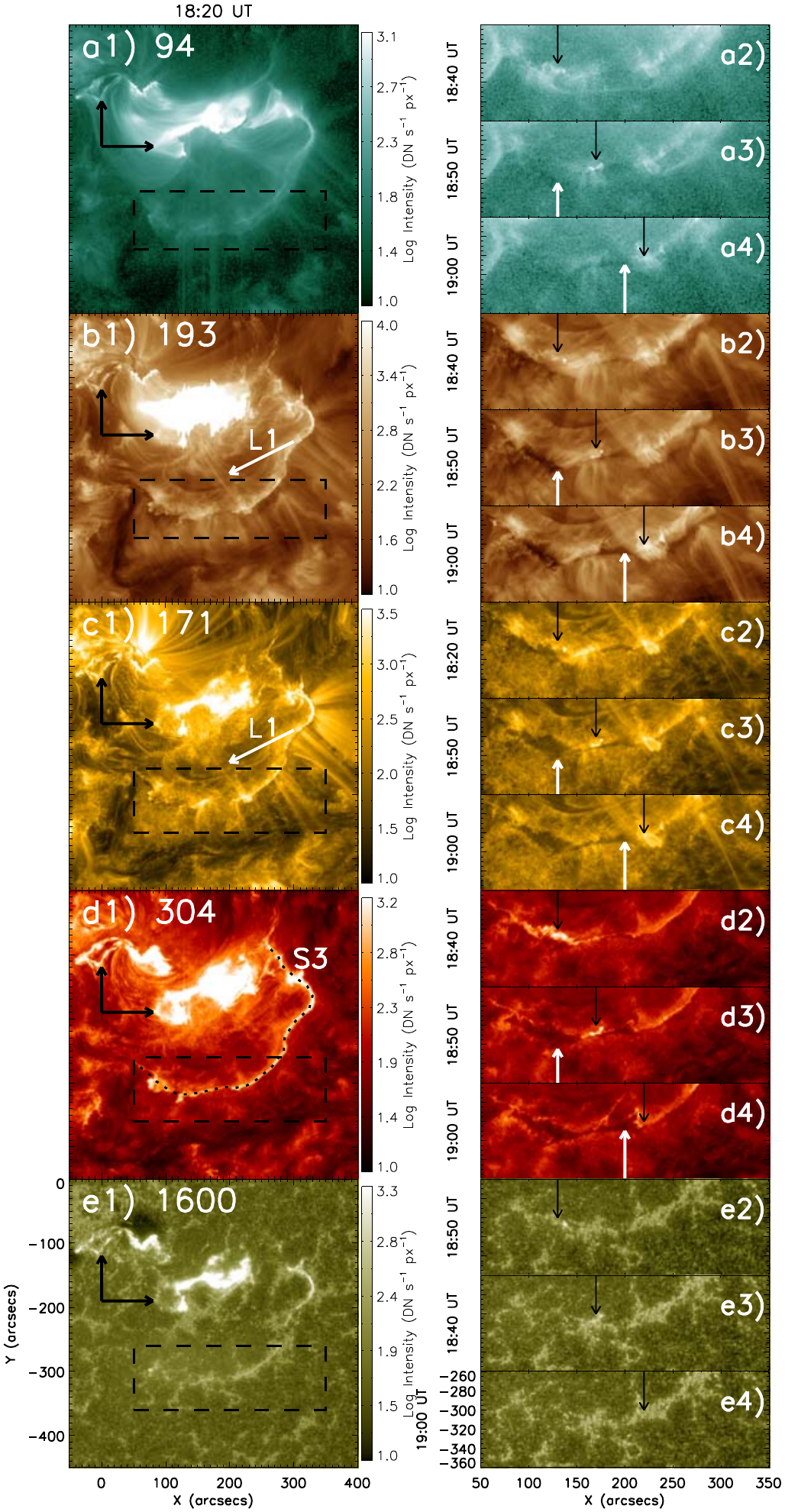}
\caption{The slipping motion along SFR in AIA 94, 193, 171, 304, and 1600~{\AA}. In the left column, the white arrows show the survived warm coronal loop connecting SFR to N2, and the black arrows indicate the hooks of FRs. In the right column, the black arrows point out the moving bright features, and the white arrows in panels a-d present the appearance of the dimmings. The dashed box in panels a1-e1 illustrates the FOV of the panels of the right column, and the dotted line in panel d1 is used to obtain the time-slice images in Figure 4b-c.
\label{f3}}
\end{figure}

\clearpage

\begin{figure}
\epsscale{0.9}
\plotone{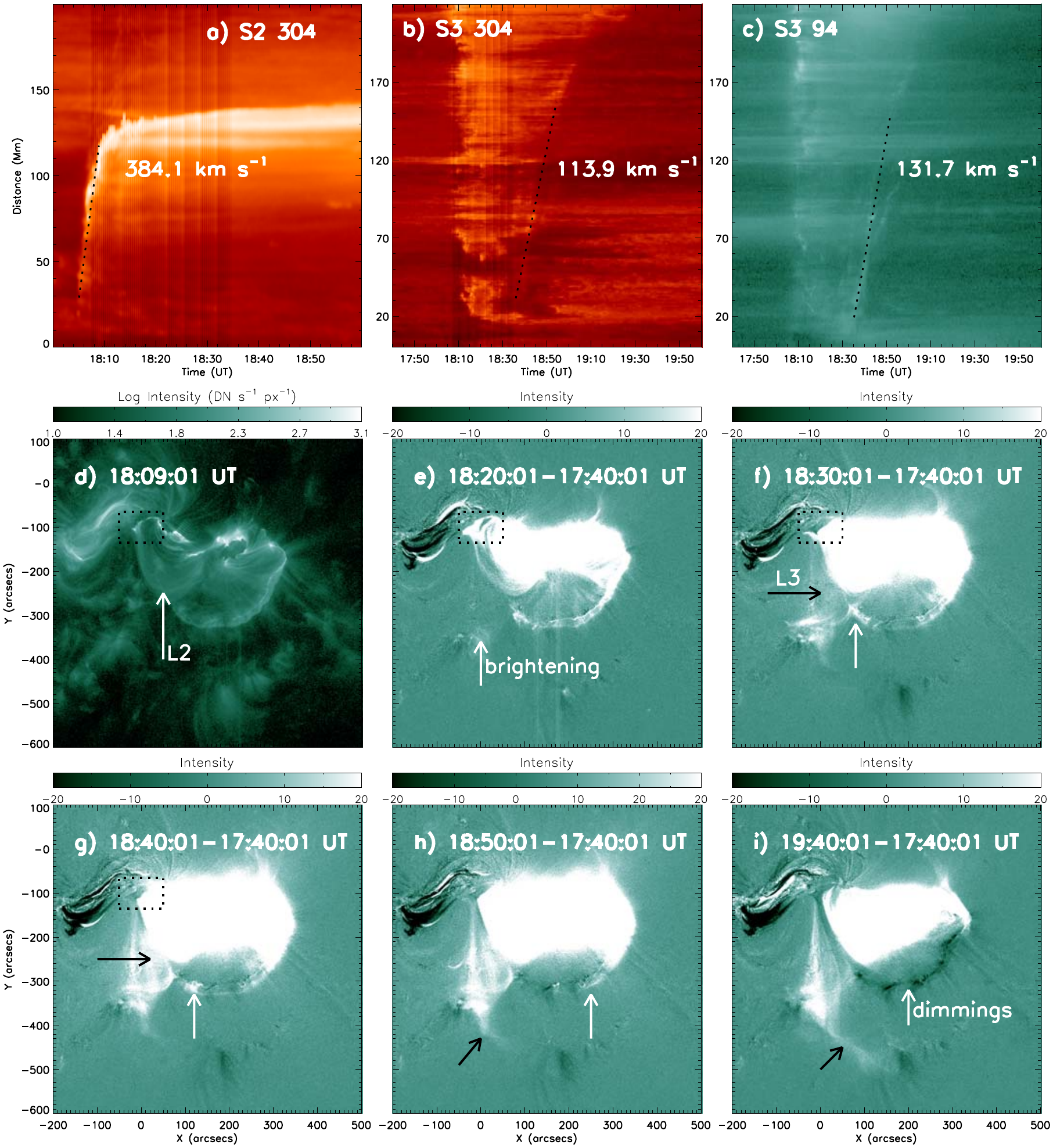}
\caption{The slipping motions of FR1 (panel a) and SFR (panels b-c) in time-slice images along the dotted lines in Figure 2 and 3, and the flare evolution in original (panel d) and bas-difference (panels e-i) images of AIA 94~{\AA}. The linear fitted velocities are obtained along the dotted lines. The black arrows in panels f-g show the newly-formed L3, and the white arrows in panels e, f-h, and i indicate the brightening, the moving features, and the dimmings, respectively. The dotted boxes in panels d-g illustrate the fading of the hook of FR2.
\label{f4}}
\end{figure}

\clearpage

\begin{figure}
\epsscale{0.9}
\plotone{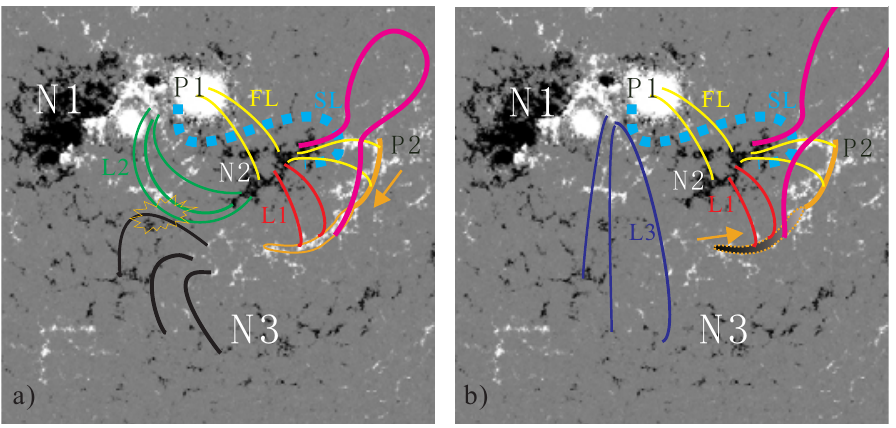}
\caption{The sketch showing the SFR's slippage and the formation of L3, based on the HMI magnetogram showing the associated magnetic polarities (P1-P2, N1-N3). The colourful lines indicate SL (light-blue), FLs (yellow), L1 (red), L2 (green), L3 (deep-blue), the erupting structure (pink), and FR3-SFR (orange). The shadow within the hook of SFR represents coronal dimmings, and the orange arrows denote the direction of the slipping motion along the outer and inner edge of SFR. The black lines show the overlying loops of F3, and the sparkle represents the reconnection.
\label{f5}}
\end{figure}

\end{document}